\documentclass[useAMS,usenatbib]{mn2e}
\usepackage{epsfig}
\usepackage{color}
\usepackage{subfigure}

\def\lsim{\mathrel{\rlap{\lower3.5pt\hbox{\hskip0.5pt$\sim$}}
    \raise0.5pt\hbox{$<$}}}
\def\gsim{~\rlap{$>$}{\lower 1.0ex\hbox{$\sim$}}}

\newcommand{\goodgap}{\hspace{\subfigtopskip} \hspace{\subfigbottomskip}}

\title[Spherical galaxy models and Tsallis statistics]{Spherical galaxy models as equilibrium configurations in nonextensive statistics}
\author[V.F. Cardone et al.]{V.F. Cardone$^{1,2}$, M.P. Leubner$^{3}$, A. Del Popolo$^{4}$\\
$^1$I.N.A.F. - Osservatorio Astronomico di Roma, via Frascati 33, 00040\,-\,Monte Porzio Catone (Roma), Italy \\
$^2$Dipartimento di Scienze Fisiche, Universit\`{a} degli Studi di Napoli "Federico II",
Complesso Universitario \\ di Monte Sant'Angelo, Edificio N, via Cinthia, 80126 - Napoli, Italy \\
$^3$Institut f$\ddot{u}$r Astro\,- und Teilchenphysik, Leopold\,-\,Franzes\,-\,Universit$\ddot{a}$t Innsbruck, Technikerstr., 25, 6020, Innsbruck, Austria \\
$^4$Dipartimento di Fisica e Astronomia, Universit\`{a} di Catania, Viale Andrea Doria 6, 95125 - Catania, Italy}

\date{Accepted xxx, Received yyy, in original form zzz}

\begin{document}

\maketitle

\begin{abstract}

Considering galaxies as self\,-\,gravitating systems of many collisionless particles allows to use methods of statistical mechanics inferring the distribution function of these stellar systems. Actually, the long range nature of the gravitational force contrasts with the underlying assumptions of Boltzmann statistics where the interactions among particles are assumed to be short ranged. A particular generalization of the classical Boltzmann formalism is available within the nonextensive context of {\it Tsallis $q$\,-\,statistics}, subject to non\,-\,additivity of the entropies of sub\,-\,systems. Assuming stationarity and isotropy in the velocity space, it is possible solving the generalized collsionless Boltzmann equation to derive the galaxy distribution function and density profile. We present a particular set of nonextensive models and investigate their dynamical and observable properties. As a test of the viability of this generalized context, we fit the rotation curve of M33 showing that the proposed approach leads to dark matter haloes in excellent agreement with the observed data.

\end{abstract}

\begin{keywords}
galaxies\,: kinematic and dynamics -- dark matter -- galaxies\,: spiral
\end{keywords}

\section{Introduction}

Consider a galaxy as an ensemble of many particles evolving under the action of their own gravitational potential. The enormous number of constituents ($\sim 10^{9}$\,-\,$10^{11}$) prevents the solution of the $N$\,-\,body problem both, analytically and numerically. Hence, along the history of galaxy dynamics studies, different methods have been developed to find a way modeling density profiles of galaxies and derive their kinematic and observable properties. Notwithstanding the fine details, such techniques may be divided into three broad classes.

First, one can rely on an empirical approach starting from observations and deducing a theoretical model. For instance, photometric measurements allow to track the galaxies surface brightness profile, which can be deprojected under suitable assumptions, yielding the three dimensional luminosity density. Subsequently, kinematic measurements or stellar population synthesis codes allow to determine the actual ingredients (e.g., the stellar mass\,-\,to\,-\,light ratio) leading finally to a particular mass model. Typical examples of this approach are the exponential model \citep{F70} for spiral galaxies discs and the PS \citep{PS97} profile for the luminous component of early\,-\,type galaxies.

Based on observations, these methods certainly cannot be used to invent models for dark haloes, galaxies are supposed to be embedded in. This naive consideration has motivated the development of increasingly high accurate numerical N\,-\,body simulations, which follow the evolution of many particles taking care of the nonlinearities induced by the gravitational interactions of each particle with the rest of the system. Based on the analysis of the results of these simulations, a plethora of models for the dark matter component has been proposed along the years, typically subject to a double power\,-\,law profile, i.e. $\rho \propto (r/r_s)^{-\alpha} (1 + r/r_s)^{-(3 - \alpha)}$. While there is a general consensus that the density scales as $r^{-3}$ for large $r$, there is still an open debate on the value of $\alpha$, adopting $\alpha = 1$ for the popular NFW model (Navarro et al. 1996, 1997) or $\alpha = -1.5$ \cite{Moore+98,Ghigna+00} for a steeper and $\alpha \simeq 0.7$ \citep{Power+03} for a shallower model. It is also possible that the logarithmic density slope never attains a finite asymptotic value in the centre \citep{Navarro+04} thus favouring exponential\,-\,like models as the Einasto profile \citep{Ein65,Polls}. The still open debate on both, the shape of the density profile and its universality along with the disagreement regarding the rotation curves of low mass systems (see, e.g., de Blok 2010 and refs. therein) are strong evidences that something is still missing or not well understood in the N\,-\,body simulations, such that the resulting density profiles should rather be taken {\it cum grano salis}.

As third approach, we may interpret the observations in view of many particle systems, where stars in a galaxy are similar to the molecules in a gas. One can therefore rely on the methods of statistical mechanics to infer the distribution function (DF) of galaxies which, once integrated over the velocity space, gives the mass density profile. The context of standard Boltzmann-Gibbs-Shannon (BGS) statistics along with a DF depending on the total energy only leads to the conventional isothermal sphere model as equilibrium configuration. However, there is a fundamental difference between stars in a galaxy governed by long-range gravitational interactions and the short range correlations between molecules in a gas. Consequently, BGS statistics cannot be applied rigorously although it is commonly assumed that the effect of this systematic error is negligible. A generalization of the BGS statistics has been proposed dealing with systems subject to long-range interactions besides other nonlinear phenomena. Tsallis (1988) first proposed the generalized entropy functional of a system as

\begin{equation}
S_q = -k \sum{\frac{p_i^{q} (p_i^{1 + q} - 1)}{1 - q}}
\end{equation}
where $k$ is the Boltzmann constant, $p_i$ the probability of the $i$\,-\,th microstate and the entropic index $q$ is a real number whose deviation from unity parameterizes the departure from the standard BGS entropy where

\begin{equation}
\lim_{q \rightarrow 1}{S_q} = - k \sum{p_i \ln{p_i}}
\end{equation}
recovers the classical entropy function. The most distinguishing feature of the generalized nonextensive entropy is manifest in the pseudo-additivity as\,:

\begin{equation}
S_q(A + B) = S_q(A) + S_q(B) + (1 - q) k^{-1} S_q(A) S_q(B)
\end{equation}
which again reduces to the standard BGS additivity for $q = 1$. It is worth stressing that the nonlinear term on the right hand side accounts for long-range interactions coupling two spatially separated systems. Today, nonextensive statistics in the context of the Tsallis entropy generlizations is remarkably successful and widely applied in a variety of different research fields.

A first application in galactic dynamics has been performed by maximizing the Tsallis entropy for fixed total mass and energy thus obtaining the analog of polytropic models for stellar systems \citep{PP93,TS02}. Later, Leubner (2005) considered the generalized Boltzmann equation and derived density profiles for the dark matter haloes in good agreement with N\,-\,body simulations \citep{K06}. In a first approximation Leubner (2005) assumed that the entropic index $q$ is spatially constant in the system. Indeed, since $q \ne 1$ describes the deviations from classical statistics due to gravitational interactions, it is expected that $q$ significantly differs from 1 where the gravitational potential is strong, but recovering the classical value $q = 1$ for vanishing potential in the outskirts. Later, a fundamental relation combining the spatial dependence of the entropic index $q$ with the density and the velocity dispersion was provided \citep{Du07}, without performing a solutions to the general problem. Based on this previous analysis, we make a step further by investigating the kinematic and observable properties of a class of models obtained by searching for equilibrium configurations in the framework of nonextensive statistics, assuming spherical symmetry but keeping the spatial dependence of the entropic index $q$ along with a suitable ansatz for the velocity dispersion required to close the system of equations.

In Sect.\,2, we review the derivation of the set of equations defining equilibrium configurations in generalized $q$\,-\,statistics. After introducing in Sect.\,3 a suitable empirical form for the velocity dispersion, we also investigate the basic kinematics and observable quantities as function of the model parameters. As a test case, we fit the rotation curve of M33 using the nonextensive approach for the dark halo component. Finally, a summary of results is provided in Sect.\,4.

\section{Equilibrium configurations}

In the context of theoretical galactic dynamics a system is fully characterized by the DF defining the number density of stars in the configuration space. Hence, $f({\bf r}, {\bf v}, t)$ is the number of particles which, at a given time $t$, have a position in an infinitesimal cube $d{\bf r}$ centred on ${\bf r}$ and a velocity in a cube $d{\bf v}$ centred on ${\bf v}$. Adopting BGS statistics, the DF must be a solution of Boltzmann's equation\,:

\begin{equation}
\frac{\partial f}{\partial t} + {\bf v} \cdot \frac{\partial f}{\partial {\bf r}} - \nabla \phi \cdot \frac{\partial f}{\partial {\bf v}} = {\cal{C}}(f) \ ,
\label{eq: boltz}
\end{equation}
where $\phi$ constitutes the gravitational potential and ${\cal{C}}(f)$ is the collision term. Assuming a stationary and collisionless system, the solution of Eq.(\ref{eq: boltz}) for spherical symmetry yields the isothermal configuration as \citep{BT87}\,:

\begin{equation}
f({\bf r}, {\bf v}) = \frac{\rho_s}{(2 \pi \sigma^2)^{3/2}} \exp{\left [ \frac{{\bf v^2}/2 - \psi(r)}{\sigma^2} \right ]}
\label{eq: isodf}
\end{equation}
where $\rho_s$ denotes the density at the centre, $\sigma$ is the (constant) velocity dispersion and $\phi(r) = -\phi(r) + \phi_0$ represents the relative potential. Integrating over the velocity space yields the standard isothermal sphere solution as BGS equilibrium configuration.

For the nonextensive generalization we follow Du (2007) defining the $q$\-\,logarithmic and exponential functions as\,:

\begin{displaymath}
\ln_q{f} = \frac{f^{1 - q} - 1}{1 - q} \ \ , \ \ e_q(f) = [1 + (1 - q) f]^{1-q} \ \ ,
\end{displaymath}
which reduce to the classical natural logarithm and exponential functions in the limit $q \rightarrow 1$ (i.e., when Tsallis statitstics reduces to the BGS case). The generalized Boltzmann equation has the same formal content as (\ref{eq: boltz}), but the collision term is now evaluated in a different way (see Lima et al. 2001 and Du 2007 for the full expression). Solving the corresponding Lagrangian variational problem for the equilibrium configuration, the resulting DF reads\,:

\begin{equation}
f_q(\varepsilon_q) = A_q [1 - (1 - q) \beta \varepsilon_q]^{1/(1 - q)}
\label{eq: fq}
\end{equation}
with $\beta$ being a Lagrange parameter and

\begin{equation}
\varepsilon_q = \frac{\left \{ 1 - \left [ 1 - (1 - q) \beta m v^2/2 \right ]
[1 + (1 - q) \beta m \psi] \right \}}{(1 - q) \beta}
\label{eq: energy}
\end{equation}
represents the sum of the potential and kinetic energy of the system. Note that the total energy is nonextensive accounting for long-range interactions and correlations in the system. Eq.(\ref{eq: fq}) can be conveniently rewritten in terms of the usual variables introducing the velocity dispersion $\sigma$ as\,:

\begin{eqnarray}
f_q({\bf r}, {\bf v}) & = & \frac{B_q \rho_s}{(2 \pi \sigma^2)^{3/2}} \left [ 1 + (1 - q) \psi/\sigma^2 \right ]^{\frac{1}{1 - q}} \nonumber \\
~ & \times & \left [ 1 - (1 - q) v^2/2\sigma^2 \right ]^{\frac{1}{1 - q}} \ ,
\label{eq: fqdf}
\end{eqnarray}
where now $\sigma = \sigma(r)$ is a function of the radial coordinate and $B_q$ is a normalization constant. For $q \rightarrow 1$, $f_q$ reduces to the Maxwellian DF and Eq.(\ref{eq: fqdf}) denoted the nonextensive generalization. Integrating over the velocity space the density is obtained as\,:

\begin{equation}
\rho = \rho_s \left [ 1 + (1 - q) \psi/\sigma^2 \right ]^{\frac{1}{1 - q}}
\label{eq: rhoq}
\end{equation}
where the potential $\psi$ is defined by Poisson's equation for spherical symmety\,:

\begin{equation}
\frac{1}{r^2} \frac{d}{dr} \left ( r^2 \frac{d\psi}{dr} \right ) = - 4 \pi G \rho \ .
\label{eq: poisson}
\end{equation}
On the other hand, considering Eq.(\ref{eq: fqdf}) for an equilibrium configuration, Boltzmann's equation (\ref{eq: boltz}) for a collisionless system must be solved leading to a relation between the entropic index $q$ and the velocity dispersion as \citep{Du07}\,:

\begin{equation}
1 - q = -2 \sigma \left ( \frac{d\sigma}{dr} \right ) \left ( \frac{d\phi}{dr} \right ) = - 2 \sigma \left ( \frac{d\sigma}{dr} \right )
\left [ \frac{G M(r)}{r^2} \right ]^{-1} \ ,
\label{eq: qsigma}
\end{equation}
which leads after combining with Eq. (\ref{eq: poisson}) to\,:

\begin{equation}
1 - q =   - \frac{\sigma \nabla^2 \sigma + (\nabla \sigma)^2}{2 \pi G \rho} \ .
\label{eq: qsigmaeqend}
\end{equation}
Finally, inserting Eq.(\ref{eq: rhoq}) into Eq.(\ref{eq: poisson}) and eliminating the potential $\psi$, yields a second order differential equation for the density profile\,:

\begin{equation}
\frac{d^2 \rho}{dr^2} + \frac{2}{r} \frac{d\rho}{dr} - \frac{q}{\rho} \left ( \frac{d\rho}{dr} \right )^2
+ \frac{2}{\sigma} \left ( \frac{d\sigma}{dr} \right ) \left ( \frac{d\rho}{dr} \right ) + \frac{4 \pi G \rho^{q + 1}}{\sigma^2 \rho_s^{q - 1}} = 0 \ .
\label{eq: rhoeqend}
\end{equation}
Eqs.(\ref{eq: qsigmaeqend}) and (\ref{eq: rhoeqend}) represent a system of two coupled nonlinear differential equations for the the radial dependence of three variables, the density $\rho(r)$, the velocity dispersion $\sigma(r)$ along with the  entropic index $q(r)$. When $\sigma = const$, Eqs.(\ref{eq: qsigmaeqend}) and (\ref{eq: rhoeqend}) reduce to those provided by Leubner (2005), while, in the limit $q \rightarrow 1$, Eq.(\ref{eq: rhoeqend}) reduces to the BGS isothermal sphere condition with $\sigma = const$ such that Eq.(\ref{eq: qsigmaeqend}) becomes an identity.

\section{Spherical models}

Derived only under the assumption of spherical symmetry, Eqs.(\ref{eq: qsigmaeqend}) and (\ref{eq: rhoeqend}) describe the equilibrium configuration of a spherically symmetric gravitational system of collisionless particles in the framework of Tsallis $q$\,-\,statistics. The system is not closed since we have only two equations to determine the three quantities $\rho(r), \sigma(r)$ and $q(r)$. In order to find a solution, we have to make an ansatz for one of these quantities and check a posteriori whether the corresponding density profile is physically reasonable. For instance, we may postulate a radial dependence of the entropic index $q(r)$, rather unsuitable since this parameter is related to complex nonlinear interaction phenomena. Alternatively, it is easier to parameterize in a model independent way the form of the velocity dispersion looking at the measured profiles in literature. We assume the following expression for the velocity dispersion \citep{N10}\,:

\begin{equation}
\sigma(r) = \sigma_0 \left [ 1 + \left ( \frac{r}{r + r_0} \right )^{\eta} \right ]^{-1}
\label{eq: sigmamodel}
\end{equation}
where $(r_0, \sigma_0)$ are scaling quantities and $\eta$ controls the transition from the varying behaviour to the constant asymptotic value. A caveat is in order here. Eq.(\ref{eq: sigmamodel}) was proposed to describe the radial component $\sigma_r$ of the velocity dispersion since, when projected along the line of sight (los), it assumes a constant $\sigma_{los}$ profile in agreement with planetary nebulae measurements in NGC\,4374. We have here taken this same phenomenological expression as an input for the $\sigma(r)$ function entering the DF, interpreting $\sigma(r)$ as a velocity dispersion. Actually, such an identification is only a formal one, but what physically $\sigma(r)$ represents is not fully understood. To understand this point, let us consider the limit $q = 1$ so that the DF reduces to the Maxwellian one. This gives rise to the well known isothermal sphere model having the remarkable property that the radial and the tangential velocity dispersion are equal. In such a case, $\sigma(r)$ turns out to be both the total velocity dispersion and the radial one (modulo a scaling factor). On the contrary, when $q \neq 1$, it is not clear whether $\sigma(r)$ should be interpreted as the radial $\sigma_r(r)$ or the total velocity dispersion or as a fully different quantity having the same physical dimensions as $\sigma_r(r)$. Should one interpret $\sigma(r)$ in Eq.(\ref{eq: fqdf}) as the radial velocity dispersion $\sigma_r(r)$, then one could rely on the spherical Jeans equation relating $\sigma_r(r)$ and $\rho(r)$ and assume a parametrized expression for the anisotropy profile $\beta(r) = 1 - \sigma_{\theta}(r)/\sigma_r(r)$. Adding this relation to Eqs.(\ref{eq: qsigmaeqend}) and (\ref{eq: rhoeqend}), a closed system for the three unknowns $\rho(r)$, $\sigma(r)$, $q(r)$ is obtained and can be solved numerically. Although such an approach is worth to be explored, it is nevertheless flawed by the theoretical uncertainty on validity of the identification $\sigma(r) = \sigma_r(r)$ for non Maxwellian DFs and still need an assumption for a function, namely the anisotropy profile. We therefore prefer to adopt a phenomenological parameterization of $\sigma(r)$ to avoid giving a conclusive interpretation of its physical meaning\footnote{A similar situation also takes place when one tries to model the coarse grained distribution function, defined as $\rho(r)/\sigma^3(r)$. Depending on whether $\sigma(r)$ is meant as the total or the radial velocity dispersion, one gets different results for the density profile and the other quantities of interest.}.

Defining the following dimensionless quantities\,:

\begin{figure*}
\centering
\subfigure{\includegraphics[width=7.5cm]{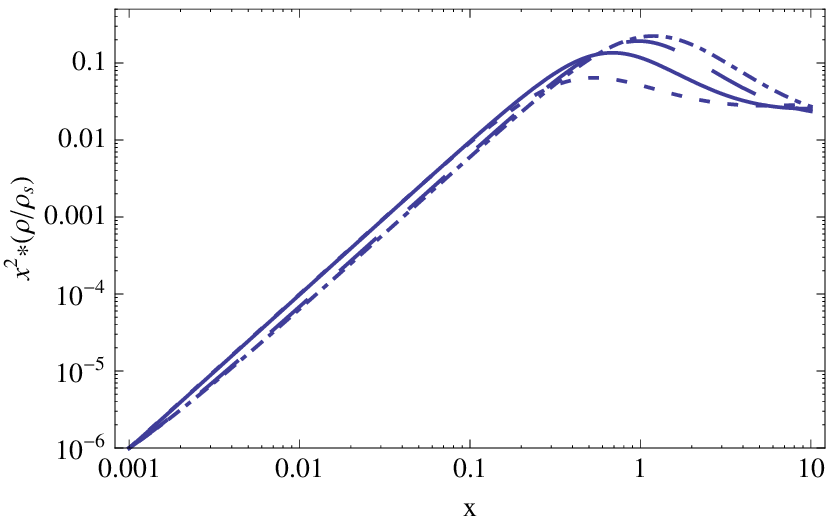}} \goodgap
\subfigure{\includegraphics[width=7.5cm]{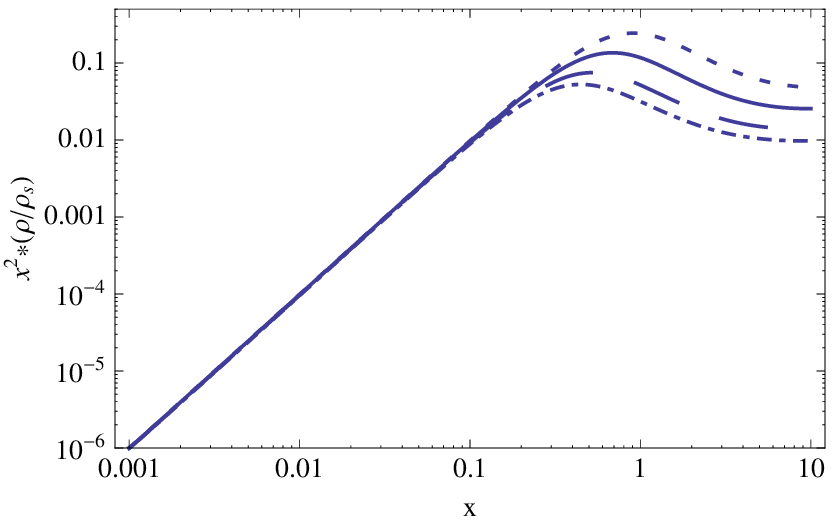}} \goodgap \\
\caption{Scaled density profile $\tilde{\rho} = \rho/\rho_s$ as a function of the dimensionless radius $x = r/r_0$ for models with $\alpha_0 = -0.5$ and different values of $(\eta, {\cal{K}})$. In the left panel, we set ${\cal{K}} = 15.0$ and $\eta = 0.1$ (short dashed), $1.0$ (solid), $2.0$ (long dashed), $3.0$ (dot dashed). Similarly, in the right panel, $\eta = 1.0$ and we consider models with ${\cal{K}} = 7.5$ (short dashed), $15.0$ (solid), $30.0$ (long dashed), $45.0$ (dot dashed).}
\label{fig: rhofigs}
\end{figure*}

\begin{displaymath}
x \equiv r/r_0 \ \ , \ \ \tilde{\rho} \equiv \rho/\rho_s \ \ , \ \ \tilde{\sigma} \equiv \sigma/\sigma_0 \ \ ,
\end{displaymath}
we can conveniently write the system (\ref{eq: qsigmaeqend})\,-\,(\ref{eq: rhoeqend}) as\,:

\begin{equation}
\frac{d^2 \tilde{\rho}}{dx^2} + \left [ (2/x) + {\cal{S}}_{\rho}(x, \eta) \right ] \frac{d\tilde{\rho}}{dx} - \frac{q}{\tilde{\rho}}
\left ( \frac{d\tilde{\rho}}{dx} \right )^2 + {\cal{K}} \frac{\tilde{\rho}^{q + 1}}{\tilde{\sigma}^2} = 0 \ ,
\label{eq: rhoeqscaled}
\end{equation}

\begin{equation}
1 - q = \frac{2/{\cal{K}}}{\tilde{\rho}} \cdot {\cal{S}}_{q}(x, \eta) \ ,
\label{eq: qeqscaled}
\end{equation}
with\,:

\begin{equation}
{\cal{K}} = \frac{4 \pi G \rho_s r_0^2}{\sigma_0^2} \ ,
\label{eq: defkappa}
\end{equation}

\begin{eqnarray}
{\cal{S}}_{\rho}(x, \eta) & = & \frac{2}{\tilde{\sigma}} \frac{d\tilde{\sigma}}{dx} \\
~ & = & - \frac{2 \eta x^{-(1 - \eta)} (1 + x)^{-(1 + \eta)}}{1 + [x/(1 + x)]^{\eta}} \ , \nonumber
\label{eq: defsrho}
\end{eqnarray}

\begin{eqnarray}
{\cal{S}}_{q}(x, \eta) & = & - [\sigma \nabla^2 \sigma + (\nabla \sigma)^2] \\
~ & = & \frac{\{ 2 [x/(1 + x)]^{\eta} - 1 \} \eta^2 - \{1 + [x/(1 + x)]^{\eta} \} \eta}
{x^{2 - \eta} (1 + x)^{2 + \eta} \{ 1 + [x/(1 + x)]^{\eta} \}^4} \ . \nonumber
\label{eq: defsq}
\end{eqnarray}
Eq.(\ref{eq: qeqscaled}) is now a simple algebraic equation for $q(\eta)$ to be solved trivially and inserted into Eq.(\ref{eq: rhoeqscaled}). This procedure generates a complex second order nonlinear differential equation for the scaled density $\tilde{\rho}$, which can be straightforwardly integrated numerically provided that initial conditions are given as follows\,:

\begin{displaymath}
\tilde{\rho}(x_{min}) = 1 \ \ , \ \ \tilde{\rho}^{\prime}(x_{min}) = \alpha_0/x_{min} \ \ ,
\end{displaymath}
where $x_{min}$ is a very small but finite number, the prime denotes the derivative with respect to $x$ and $\alpha_0 = d\ln{\tilde{\rho}}/d\ln{x}(x = x_{min})$ is the logarithmic density slope in $x_{min}$. Note that one should in principle set $x_{min} = 0$ since, by definition, $\tilde{\rho}(0) = 1$. However, because of numerical problems, we will never solve Eq.(\ref{eq: rhoeqscaled}) up to $x_{min} = 0$, but rather set $x_{min} = 10^{-5}$ and assume that the difference $\rho(x_{min}) - \rho(x = 0)$ is negligible. As a further remark, we note that $\alpha_0$ must be zero to assure that the density does not diverge at the centre and $\rho_s$ in the DF (\ref{eq: fqdf}) takes a finite value everywhere. However, this does not prevent us from taking a negative value for $\alpha_0$ if it is evaluated in $x_{min} > 0$, still imposing $\alpha(x = 0) = 0$. While this introduces a non-realistic abrupt change in the logarithmic density slope profile, we nevertheless prefer leaving $\alpha_0$ as a free parameter in order to explore a larger class of models. Thus, the resulting models are not suitable for $x \le x_{min}$, but $x_{min}$ is small enough to avoid any impact on astrophysical application of interest. With this caveat in mind, we explore the physical properties of the model obtained by solving Eqs.(\ref{eq: rhoeqscaled})\,-\,(\ref{eq: qeqscaled}) as a function of the three parameters $(\eta, {\cal{K}}, \alpha_0)$, while stressing that $(r_0, \sigma_0)$ are only scaling parameters.

\begin{figure*}
\centering
\subfigure{\includegraphics[width=7.5cm]{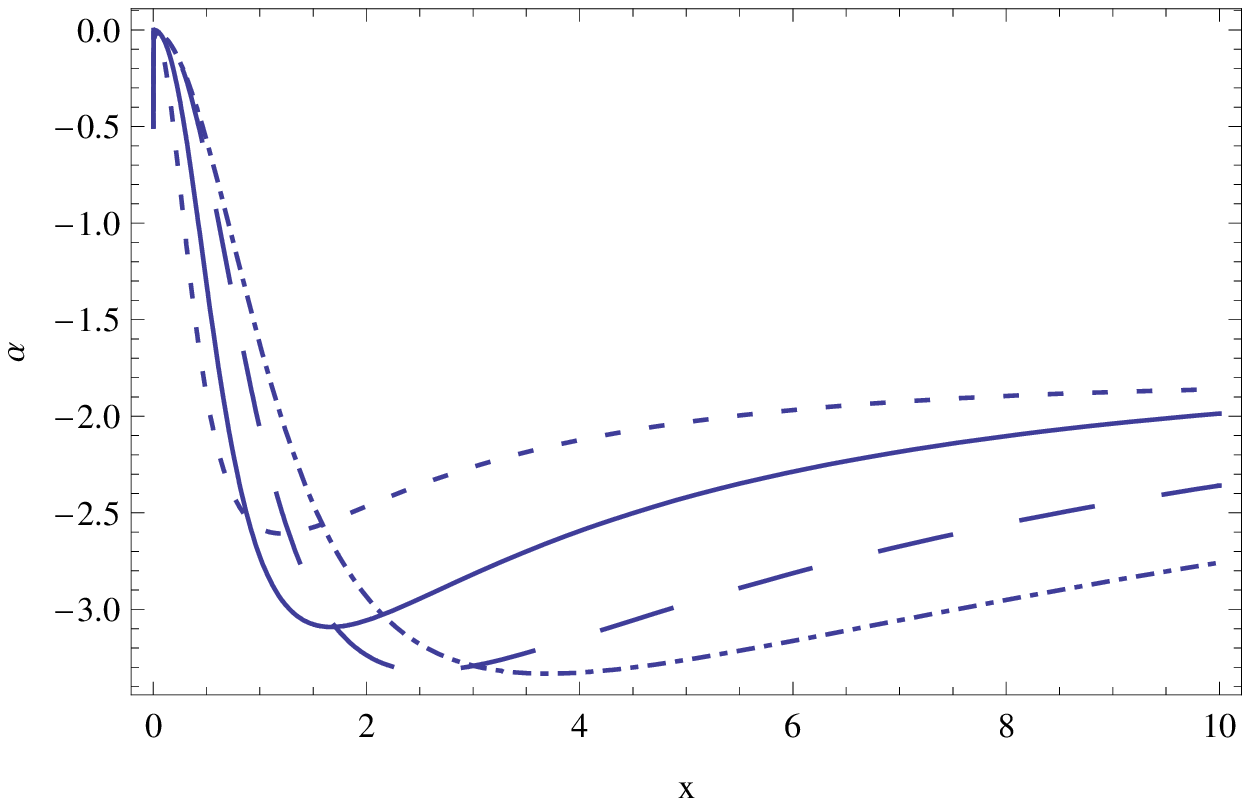}} \goodgap
\subfigure{\includegraphics[width=7.5cm]{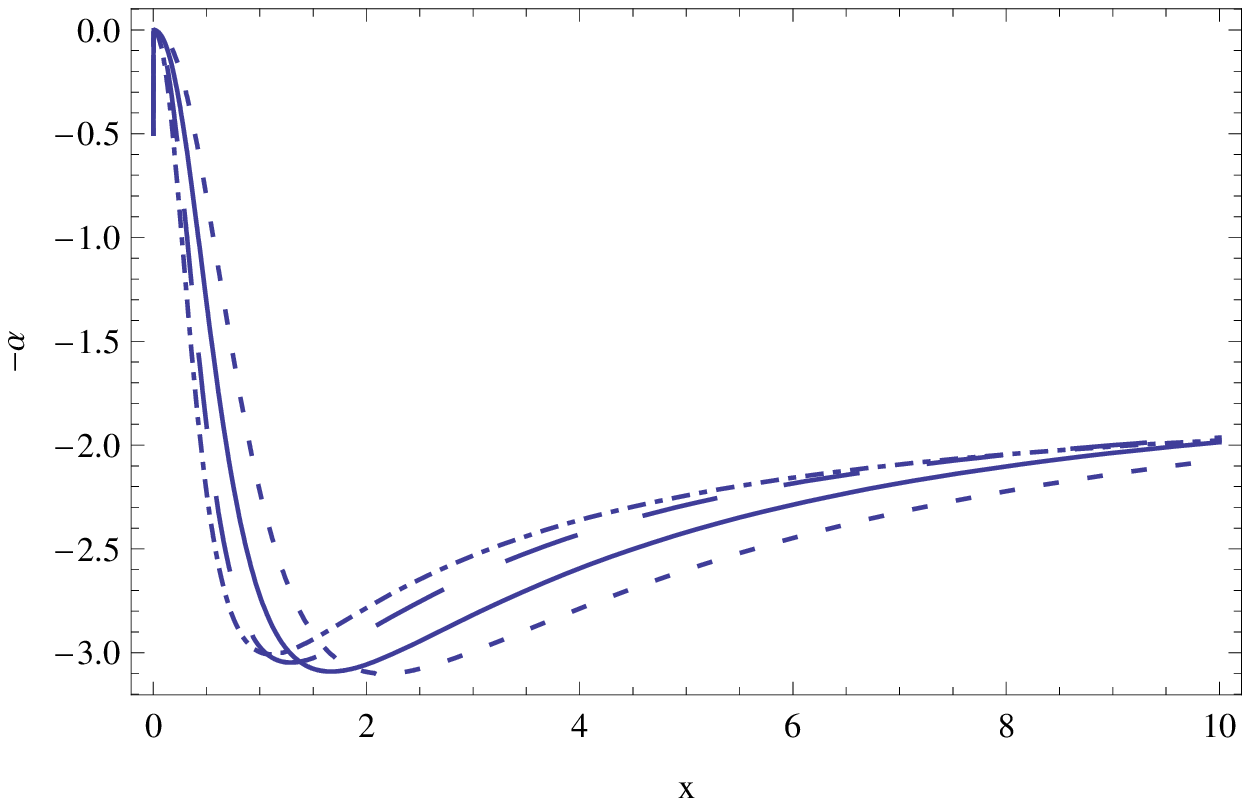}} \goodgap \\
\caption{Same as Fig.\,\ref{fig: rhofigs} but for the logarithmic density slope $\alpha = d\ln{\tilde{\rho}}/d\ln{x}$.}
\label{fig: alphafigs}
\end{figure*}

Before discussing the properties of the model thus obtained, it is worth pointing at an underlying assumption. As Eq.(\ref{eq: fqdf}) shows, the DF is isotropic in the velocity space so that the velocity distribution function $\phi({\bf v})$ (i.e., the integral of the DF over the position coordinates) is the same along both the radial and tangential direction. Actually, recent numerical simulations \citep{H06,H09,K10,L10} have pointed out at a marked difference. To overcome this issue, one should modify the dependence of the DF on the velocity which now enters only through its magnitude (entering the total energy), but not its separated components. In such a case, however, the DF would not be the solution of the variational problems for the non extensive statistics which should then be reformulated from the beginning. In this first step analysis, we therefore to retain the assumption $\phi(v_r) = \phi(v_t)$ warning the reader that the resulting model should be used only for those subset of dark matter haloes where this ans\"atz approximately holds true.

\subsection{Density profile}

In order to solve numerically Eq.(\ref{eq: rhoeqscaled}), we need to provide the parameters $(\eta, {\cal{K}}, \alpha_0)$ along with the initial conditions and choose a range for the scaled radius $x$. We fix $(x_{min}, x_{max}) = (10^{-5}, 10)$ such that the numerical solution can be reliably performed avoiding both, the very inner parts affected by convergence problems and the very outer regions where a significant loss of precision makes the numerical solution unstable. As we will see later, the spatial range adopted is much wider than the one usually probed by observations, wherefore no region of interest is excluded.

Fig.\,\ref{fig: rhofigs} shows the scaled density $\tilde{\rho} = \rho/\rho_s$ as function of the dimensionless radius $x = r/r_0$ for different choices of the model parameters\footnote{The motivations for the values of $(\eta, {\cal{K}}, \alpha_0)$ adopted in this and the following figures will be discussed later .} $(\eta, {\cal{K}}, \alpha_0)$. As a general result, we find that the density profile decreases everywhere so that the model is physically meaningful, but the rate of this scaling with $x$ depends on the adopted parameters.

For fixed values of $(\eta, {\cal{K}})$, the density profile turns out to be almost independent of $\alpha_0$, affecting the shape of $\tilde{\rho}(x)$ only in the very inner regions $(x << 0.01)$. This is an expected result noting that $\alpha_0$ only enters into the initial conditions, thus limiting a global influence. As soon as $x$ increases, the dominant terms in Eq.(\ref{eq: rhoeqscaled}) are driven by $(\eta, {\cal{K}})$, making the exact value of $\alpha_0$ unimportant. Therefore let us assume $\alpha_0 = -0.5$ as a compromise between cored and cusped models and consider the variation of different quantities of interest as a function of $(\eta, {\cal{K}})$ only.

The density profile indeed strongly depends on these two parameters. In particular, we note that, except in the $x << 1$ limit, the larger $\eta$ is, the larger is also $\tilde{\rho}(x)$ for a fixed ${\cal{K}}$, although the different profiles converge to each other asymptotically. This effect can be qualitatively explained by looking at the assumed velocity dispersion profile in Eq.(\ref{eq: sigmamodel}). While $\sigma(r)$ always approaches a constant value, its initial decrease is more pronounced for larger $\eta$ values. Since a larger $\sigma(x)$ calls for a larger density, it is therefore clear why $\tilde{\rho}(x)$ has to be an increasing function of $\eta$ in order to reproduce the assumed $\sigma(r)$ profile. The impact of $\eta$ is, however, important only in a limited radial range and, moreover, there seems to be a sort of saturation such that values of $\eta << 1$ give rise to models that can be hardly discriminated.

${\cal{K}}$ appears as important parameter of the theory and is indirectly proportional to the density for fixed $x$ and $\eta$ values. Understanding the motivation for this behaviour is not straightforward. On one hand, Eq.(\ref{eq: defkappa}) shows that the central density $\rho_s$ is larger for larger ${\cal{K}}$ so that one naively expect $\tilde{\rho} \propto {\cal{K}}$ should Eq.({\ref{eq: defkappa}) be extrapolated to any radius. One can, however, speculate a reason for the inverse scaling of $\tilde{\rho}$ with ${\cal{K}}$ looking at Eq.(\ref{eq: qeqscaled}) for the entropic index $q(x)$. To this end, let us first note that the adopted parametrization for $\sigma(r)$ gives a velocity dispersion which quickly approaches a constant value as is the case for the isothermal sphere. As well known, this latter model is the equilibrium configuration for the BGS statistics characterized by $q = 1$. It is therefore expected that $1 - q$ is roughly constant and small everywhere but in the very inner regions where the long range gravitational interactions are maximized. As a consequence, the ratio ${\cal{S}}_\rho/({\cal{K}} \tilde{\rho})$ must be constant and, since ${\cal{S}}_{\rho}(x)$ is roughly constant for $x > 0.1 - 0.5$, so must ${\cal{K}} \tilde{\rho}(x)$ be if we want that the right hand side of Eq.(\ref{eq: qeqscaled}) has the expected behaviour. As a consequence, the larger ${\cal{K}}$ is, the smaller must be $\tilde{\rho}(x)$. We may therefore argue that the scaling of $\rho$ with ${\cal{K}}$ is just a manifestation of Tsallis statistics being a moderate correction to the BGS statistics in collisionless self\,-\,gravitating systems. Note that this is related to the particular expression adopted for $\sigma(r)$, but we can argue that a similar behaviour generally holds true provided the DF is not too different from a Gaussian.

While the density profile is always decreasing, the local slope of the $\tilde{\rho}$ vs $x$ relation is not constant. This can be clearly appreciated looking at the logarithmic density slope\,:

\begin{displaymath}
\alpha = \frac{d\ln{\rho}}{d\ln{r}} = \frac{d\ln{\tilde{\rho}}}{d\ln{x}} \ ,
\end{displaymath}
which is plotted in Fig.\,\ref{fig: alphafigs} for the same choice of model parameters as in Fig.\,\ref{fig: rhofigs}. Indeed, $\alpha$ is a strongly varying function of $x$ starting from $\alpha_0$, decreasing to $\alpha \sim -3$ for $x = x_{\alpha} \sim 1$ and then slowly approaching the isothermal $\alpha = -2$ value (outside the range plotted). It is interesting to note that it now $\eta$ playing a more important role than ${\cal{K}}$. In particular, the larger is $\eta$, the larger (in absolute value) is the minimum of $\alpha(x)$ and the later it is achieved, i.e., the larger is $x_{\alpha}$. Moreover, for $x \le x_{\alpha}$, $|\alpha(x)|$ is a quickly decreasing function of $x$, while the opposite trend is shown in the $x \ge x_{\alpha}$ range. The dependence on ${\cal{K}}$ is reversed with respect to those on $\eta$ thus mirroring what happens with $\tilde{\rho}(x)$. We can qualitatively explain the minor role of ${\cal{K}}$ noting that, according to our previous interpretation, the model evolves in a way keeping ${\cal{K}} \tilde{\rho}$ nearly constant, whatever the scaling of $\tilde{\rho}$ with $x$ is. Contrary, since $\eta$ determines the transition from the decreasing to the flat profile of $\sigma(x)$, it has to play a similar role for $\tilde{\rho}(x)$, thus explaining why $\alpha$ strongly depends on it.

The $\alpha(x)$ profile suggests that the density law could be approximated by a double power\,-\,law model such as the generalized NFW one \citep{JS00} with $\tilde{\rho} \propto (r/r_s)^{-\alpha} [1 + (r/r_s)]^{-(3 - \alpha)}$, or the $(\alpha, \beta, \gamma)$ models (Zhao 1996, 1997) having $\tilde{\rho} \propto (r/r_s)^{-\alpha} [1 + (r/r_s)^{\gamma}]^{-(\beta - \alpha)/\gamma}$. Actually, all these models present a central divergence, while our model is constructed to have $\tilde{\rho} = 1$ in $x = 10^{-5}$. Moreover, there is a single scale radius so that the transition is quite smooth, while our $\alpha(x)$ profiles show a quick transition. By trial and error, we have found that a better approximation is provided by the following profile\,:

\begin{equation}
\tilde{\rho}(x) = \left [ 1 + \left ( \frac{x}{x_c} \right )^{\alpha_{in}} \right ]^{-1} \cdot
\left [ 1 + \left ( \frac{x}{x_t} \right )^{\alpha_{t}} \right ]^{- \frac{\alpha_{out} - \alpha{in}}{\alpha_t}} \ ,
\label{eq: rhofit}
\end{equation}
yielding a small core for $x << x_c$, decreasing as $x^{-\alpha_{in}}$ for $x_c << x << x_t$ and asymptotically scaling as $x^{-\alpha_{out}}$ in the outer regions. The two radii $(x_c, x_t)$ separate the different scaling regions, while $\alpha_t$ determines the smoothness of the transition. In order to check whether Eq.(\ref{eq: rhofit}) indeed approximates well the numerical solution, we have fitted it to a large sample of models randomly generating the parameter space $(\eta, {\cal{K}}, \alpha_0)$. A satisfactory fit (with an rms percentage deviation of the order of $7\%$) is found for $\sim 50\%$ of the sample, but one should investigate in more detail which are the sets $(\eta, {\cal{K}}, \alpha_0)$ giving rise to realistic models. We therefore argue that Eq.(\ref{eq: rhofit}) provides a reasonably well fitting formula, which we checked in addition by fitting models tailored on observed galaxies. For these successful fitted models, median and median deviation of the slope parameters are found to be\,:

\begin{displaymath}
\alpha_{in} = 0.22 \pm 0.13 \ , \ \alpha_t = 3.78 \pm 1.51 \ , \ \alpha_{out} = 3.14 \pm 0.10 \ .
\end{displaymath}
We also note that about half of the successful fits are subject to very large $x_t$ values, better approximated by the simplified empirical form\,:

\begin{equation}
\tilde{\rho}(x) = \left [ 1 + \left ( \frac{x}{x_c} \right )^{\alpha_{in}} \right ]^{-\frac{\alpha_{out}}{\alpha_{in}}}
\label{eq: rhofitbis}
\end{equation}
with a single transition scale and

\begin{displaymath}
\alpha_{in} = 5.94 \pm 0.28 \ , \ \alpha_{out} = 2.41 \pm 0.08 \ .
\end{displaymath}
Investigating which model is more appropriate and how the fitting  parameters change with $(\eta, {\cal{K}}, \alpha_0)$ is outside our aims here. We nevertheless note that, whichever model is preferred, $\alpha(r/r_0 >> 1) < 10/3$ in agreement with the upper limit found by Hansen et al. (2005) for the case of constant $q$.

\subsection{Mass profile and rotation curve}

Within the restriction of spherical symmetry, the mass profile of the nonextensive model is trivially evaluated as\,:

\begin{equation}
M(x) = 4 \pi \int_{0}^{r}{r'^2 \rho(r') dr'} = 4 \pi \rho_s r_0^3 \int_{0}^{x}{x'^2 \tilde{\rho}(x')dx'} \ .
\label{eq: massdef}
\end{equation}
Fig.\,\ref{fig: massfigs} shows the scaled mass $\tilde{M}(x) = M(x)/(4 \pi \rho_s r_s^3)$ as a function of $x$ for fiducial values of $(\alpha_0, \eta) = (-0.5, 1.0)$ and different ${\cal{K}})$ (the same results hold for other values). The dependence of $\tilde{M}(x)$ on the model parameters corresponds the one of the scaled density, as expected in view of their interrelation. Note that $\tilde{M}(x)$ depends much stronger on $\eta$ than $\tilde{\rho}$ as could be anticipated, since this quantity is directly related to $\sigma(r)$, while the relation between $\sigma$ and $\tilde{\rho}$ involves an integration which smoothes out the scaling with $\eta$.

\begin{figure}
\centering
\includegraphics[width=7.5cm]{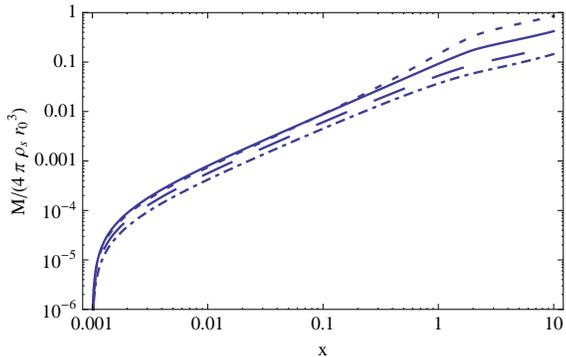}
\caption{Scaled mass $M(x)/(4 \pi \rho_s r_0^3)$ as a function of $x$ for models with $(\alpha_0, \eta) = (-0.5, 1.0)$ and ${\cal{K}} = 7.5$ (short dashed), $15.0$ (solid), $30.0$ (long dashed), $45.0$ (dot dashed).}
\label{fig: massfigs}
\end{figure}

It is worth wondering whether the total mass of the model is finite or not. On the one hand, Fig.\,\ref{fig: massfigs} shows that $\tilde{M}(x)$ keeps increasing up to the last point of the radial range probed thus suggesting an infinite total mass should the trend be extrapolated. This is consistent with $\alpha(x)$ approaching the isothermal $\alpha = -2$ value. Considering the shape of the density and logarithmic slope profiles, we argue that our model is similar to a cored isothermal sphere in the very inner and very outer regions claiming that the total mass is indeed infinite. However, Eq.(\ref{eq: rhoeqscaled}) cannot be solved to radii larger than $x_{max} = 10$ thus preventing a safe conclusion regarding the above (although reasonable) inference. Should this be the case, the model should be restricted to a maximum radius, i.e. the viral radius (defined in the way that the mean density within $R_{vir}$ is $\Delta_{vir} \bar{\rho}_M$ with $\Delta_{vir}$ depending on the cosmological model and $\bar{\rho}_M$ the mean cosmic matter density). As we will see later, $x_{vir} = R_{vir}/r_0$ is found typically well within the upper limit $x_{max} = 10$ implying securely that our solution describes the relevant quantities over the range of interest for applications. It is worth noting, however, that such a high radius cutoff is likely to be not necessary at all. Indeed, it is still possible that the infinite mass problem (if present at all and not an artifact of the numerical solver) is just a consequence of the adopted parametrization of $\sigma(r)$. Indeed, Eq.(\ref{eq: sigmamodel}) has been proposed to get an asymptotically flat dispersion at large radii just as in the case of the popular isothermal sphere. Indeed, a flat $\sigma(r)$ can alternatively be obtained postulating $M(r) \propto r$ at large radii thus giving rise to a formally infinite mass. It is therefore possible that a different choice for $\sigma(r)$, such as a double power\,-\,law, leads to a model which is similar to the one we are discussing over a large radial range, but presents a finite total mass. In order to escape the arbitrariness in the choice of a different $\sigma(r)$ profile and to avoid introducing further parameters, we will not consider this possibility, but warn the reader against concluding that non extensive statistics lead to infinite mass models.

The mass profile serves as input quantity for evaluating the rotation curve defined by $v_c^2(r) = GM(r)/r$, i.e. $v_c^2(x) \propto \tilde{M}(x)/x$. Since $\tilde{M}(x)$ approximately grows linearly for $x >> 1$, an asymptotically flat rotation curve is achieved in agreement with observations in spiral galaxies. However, the question arises whether the model we are discussing is used to describe the full mass content or the dark halo only. For a typical spiral galaxy, the stellar component is excellently described by an exponential profile, while the nonextensive approach is roughly represented by a double power\,-\,law, implying that this model is applicable for the halo component only. In this case, the entire rotation curve is obtained by summing up the stellar and dark matter contributions where the former likely dominates in the inner regions. Hence, it is important to tailor the $r_0$ value such that the regime $x >> 1$ corresponds in linear units to $r >> R_d$ with $R_d$ being the disc half mass radius. In this case a flat rotation curve can be generated in the outer regions of the disk as it is indeed observed. We will come back to this point later when considering the application to an actual case, showing how this can be easily achieved.

\begin{figure}
\centering
\includegraphics[width=7.5cm]{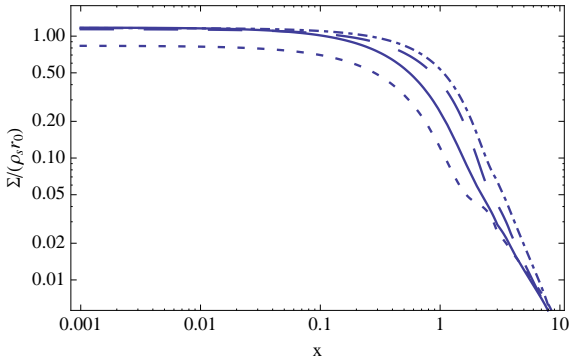}
\caption{Scaled surface density $\Sigma(x)/(\rho_s r_0)$ as a function of $x$ for models with $(\alpha_0, {\cal{K}}) = (-0.5, 15.0)$ and $\eta = 0.1$ (short dashed), $1.0$ (solid), $2.0$ (long dashed), $3.0$ (dot dashed).}
\label{fig: sigmafigs}
\end{figure}

\subsection{Surface density and projected mass}

Although argued above that our model is not suitable to describe the disc of spiral galaxies, it is nevertheless interesting to compute the surface mass density, since this quantity enters in the determination of the lensing potential and hence determines the magnification and shear induced by the lens halo \citep{SEF}. To this end, we just have to replace $x = \sqrt{\xi^2 + \zeta^2}$ in the density profile and integrate over $\zeta$, upon assuming that the line of sight coincide with the $z$\,-\,axis with $(\xi, \zeta) = (R/r_0, z/r_0)$. The results are shown in Fig.\,\ref{fig: sigmafigs} for different choices of $\eta$ and $(\alpha_0, {\cal{K}}) = (-0.5, 15.0)$ as fiducial case. As general feature we note that $\tilde{\Sigma}(x)$ first decreases approximately, changing to a shallower profile as $x$ increases. This behavior differs signifficantly from an exponential or Sersic (1968) profile wherefore it is not possible applying this model for surface brightness fits of spiral or early\,-\,type galaxies. Moreover, it enforces the previous suggestion that this model does not apply for the stellar components, but can conveniently be adopted for modelling dark haloes. Not surprizingly, concerning the dependence on the model parameters we find that the dimensionless surface density $\tilde{\Sigma}(x) = \Sigma(x)/(\rho_s r_0)$ scales with $(\eta, {\cal{K}})$ in the same way as the 3D density $\tilde{\rho}(x)$. Indeed, this is expected since the projection does not invalidate any of the argument previously discussed regarding the scaling properties of $\tilde{\rho}$.

A quantity of great interest in lensing applications is the deflection angle, which, for a spherically symmetric lens, reads $\hat{\alpha}(\xi) = M_{proj}(\xi)/(\pi \xi \Sigma_{crit})$, where

\begin{equation}
M_{proj}(\xi) = 2 \pi \int_{0}^{R}{\Sigma(R') R' dR'} = 2 \pi \rho_s r_0^3 \int_{0}^{\xi}{\tilde{\Sigma}(\xi') \xi' d\xi'}
\label{eq: defmproj}
\end{equation}
is the projected mass within the radius $R = \xi r_0$ and the critical surface density is defined as $\Sigma_{crit} = c^2 D_s/(4 \pi G D_d D_{ds})$ with $(D_d, D_s, D_{ds})$ being the observer\,-\,lens, observer\,-\,source and lens\,-\,source angular diameter distances, respectively. In case the system observer\,-\,lens\,-\,source is (nearly) perfectly aligned, the image of the lensed galaxy is a ring with radius $R_E = \hat{\alpha}(R_E)$, referred to as the Einstein radius. A measurement of $R_E$ then allows to obtain the projected mass within $R_E$ as $M_{proj}(R_E) = \pi \Sigma_{crit} R_E^2$ provided that lens and source redshifts are known and a cosmological model has been set to compute angular diameter distances.

While the dependence of $M_{proj}(\xi)$ on the model parameters is the same as for the entire mass $\tilde{M}(x)$, it is worth stressing that there is a significant difference among the various $(\eta, {\cal{K}})$ combinations. Hence, if $M_{proj}(\xi)$ is estimated from the Einstein radius in lensed systems, narrow constraints on the model parameters follow. To this end, the projected mass for $\xi > 1$ should be measured, which actually could be rather difficult. Considering, for instance, the SLACS sample \citep{Slacs}, we find $\langle R_E/R_{eff} \rangle \sim 0.6$ requiring that $r_0/R_{eff} < 1$ in order to have $M_{proj}(\xi_E)$ probing the range $\xi > 1$ where the discrimination is possible. Actually, unless the halo is quite concentrated, it is unlikely that $r_0 < R_{eff}$ leaving us in the opposite ($\xi < 1$) regime. However, we might rely on weak lensing to constrain $M_{proj}(\xi)$ at larger radii or combine lensing and central velocity dispersion measurements \citep{C09,Auger+10} for discriminating better among different $(\eta, {\cal{K}})$ values.

\begin{figure*}
\centering
\subfigure{\includegraphics[width=7.5cm]{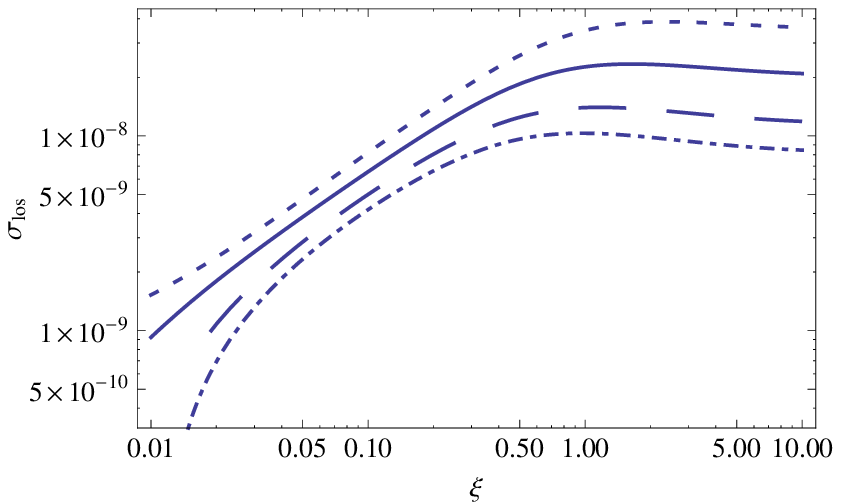}} \goodgap
\subfigure{\includegraphics[width=7.5cm]{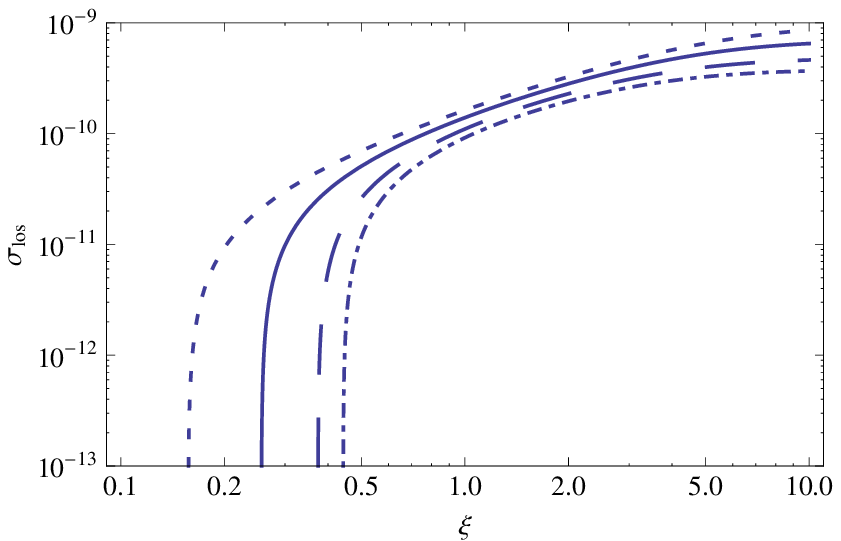}} \goodgap \\
\caption{Scaled los velocity dispersion profile assuming $r_0/R_{eff} = 1.0$ (left panel) and $10.0$ (right panel) for models with $(\alpha_0, \eta) = (-0.5, 1.0)$ and ${\cal{K}} = 7.5$ (short dashed), $15.0$ (solid), $30.0$ (long dashed), $45.0$ (dot dashed).}
\label{fig: sigmalosfigs}
\end{figure*}

\subsection{Aperture velocity dispersion}

In spiral galaxies the rotation curve is easily measured due to the presence of large amounts of gas probing the gravitational potential field. Contrary, early\,-\,type galaxies as older systems are subject to low gas content where a determination of the circular velocity is quite difficult. Moreover, these systems are typically dominated by random rather than ordered motions leading to a small $v_c(r)$, complicating quantitative measurements even more. On the other hand, the velocity dispersion profile may be traced relying on stars \citep{Cap06,Tho09} for the inner regions and, e.g., planeteary nebulae \citep{D07,Coc09} in the outskirts.

In order to determine this quantity, Jeans equations for a spherically symmetric systems must be solved first yielding the radial velocity\footnote{It is worth stressing that, should one assume that $\sigma(r)$ is yet the radial velocity dispersion, this step is redundant and one can directly project $\sigma(r)$ along the line of sight.}. Based on an ansatz for the anisotropy profile the result can be projected along the line of sight (see, e.g., \cite{BT87} for a detailed description). It is worth discussing whether such a procedure still holds in our case. Indeed, the Jeans equations follow from the moments of the collisionless Boltzmann equation indicating the implicit presence of BGS statistics. Actually, Boltzmann's equation still holds for nonextensive environments allowing to rely on the standard formalism. Hence, the $los$ velocity dispersion can still be computed as \citep{ML05}\,:

\begin{equation}
\sigma_{los}^2(R) = \frac{G M_{eff} \rho_{\star}^{eff}}{\Upsilon_{\star} I(R)} \int_{\xi_{e}}^{\infty}
{\frac{K(x_{e}/\xi_{e}) \hat{\rho}_{\star}(\xi_{e}) \hat{M}_{tot}(x_e)}{x_e} dx_e}
\label{eq: sigmalosgen}
\end{equation}
where $I(R)$ is the intensity profile, $(x_e, \xi_e) = (r/R_{eff}, R/R_{eff})$, $M_{eff}$ and $\rho_{\star}^{eff}$ are the total mass and the stellar density, respectively, at the effective radius $R_{eff}$ and the hatted quantities are normalized with respect to their values at $R_{eff}$. Finally, $M_{tot}(x_e)$ is the total mass, while $K(x_e/\xi_e)$ is a kernel function depending on the choice of the anisotropy profile. As hinted at before, in our approach, we have implicitly assumed that the radial and tangential velocity distribution functions are equal. As a consequence, our model is isotropic in the velocity space by construction so that $\beta = 0$ is the only possible choice. We take the corresponding expression for $K(x_e/\xi_e)$ from Appendix B of Mamon \& Lokas (2005) and use a de Vaucouleurs (1948) profile for the intensity law and a PS \citep{PS97} model (thus setting $\hat{\rho}_{\star} \propto x_e^{-p_n} \exp[-b_n x_e^{1/n}]$ and $n = 4$) for the stellar density.

Since we are not interested in fitting any actual dataset, the stellar component is neglected in order to demonstrate better the behavior of $\sigma_{los}$ for our model\footnote{Needless to say, this introduces a severe systematic error in the inner regions which are typically stars dominated so that the plots in Fig.\,\ref{fig: sigmalosfigs} should not be compared to measured profiles.}. Fig.\,\ref{fig: sigmalosfigs} shows some profiles for the usual combinations of model parameters and two values for the ratio $r_0/R_{eff}$ which adds now to the list of quantities $\sigma_{los}(\xi_e)$ depends on. Since $\sigma_{los}$ is related to the mass profile, it is not surprising that, for given $\xi_e$, its dependence on the $(\eta, {\cal{K}})$ parameters is the same as that of $\tilde{M}(x)$, see the discussion above. We only remark that the radial range where the $\sigma_{los}$ profiles significantly differ depends on the adopted value of the ratio $r_0/R_{eff}$. The smaller is this ratio, the larger is the radial range where discriminating among different models is possible. Moreover, $\sigma_{los}(\xi)$ is larger for smaller $r_0/R_{eff}$ value as a consequence of the interplay between the mass profile and the intensity $I(R)$. Indeed, because of the exponential term in the stellar density, the main contributions to the integral in Eq.(\ref{eq: sigmalosgen}) comes from the regions within few times $R_{eff}$ indicating that the smaller is $r_0/R_{eff}$, the more the mass term $\hat{M}(x_e)$ contributes and the larger is the los velocity dispersion.

\begin{figure}
\centering
\includegraphics[width=7.5cm]{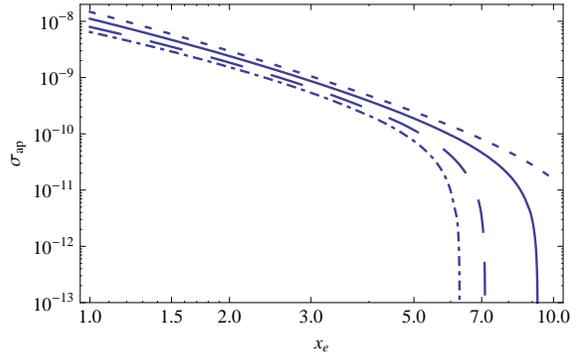}
\caption{Dimensionless aperture velocity dispersion as a function of the $r_0/R_{eff}$ ratio for a circular aperture of radius $R_{ap} = R_{eff}/2$. The model parameters are set as in Fig.\,\ref{fig: sigmalosfigs}.}
\label{fig: sigmaapfigs}
\end{figure}

Measuring $\sigma_{los}(R)$ is actually possible only if the galaxy is quite close providing spectra at different distances from the centre. For distant galaxies (for instance at intermediate redshift, typical of SLACS lenses), it is only possible to measure luminosity weighted velocity dispersion within a circular aperture of radius $R_{ap}$ set by the instrumental setup. Fig.\,\ref{fig: sigmaapfigs} presents this quantity evaluated as\,:

\begin{equation}
\sigma_{ap}(R_{ap}/R_{eff}) = \frac{2 \pi \int_{0}^{R_{ap}}{\sigma_{los}(R) I(R) R dR}}{2 \pi \int_{0}^{R_{ap}}{I(R) R dR}}
\end{equation}
and assuming $R_{ap}/R_{eff} = 1/2$, as for the median values in SLACS lenses. While the scaling with $(\eta, {\cal{K}})$ is the usual one, we note that the weighting procedure strongly smoothed out the trends, such that discriminating among different cases using $\sigma_{ap}$ only becomes very difficult unless $r_0 \sim R_{eff}$, which is quite unlikely.

\section{Testing the model}

The model we have investigated up to now is theoretically well founded and motivated by applying Tsallis statistics to the case of a collisionless self\,-\,gravitating system. However, we have to proof that this approach is also reliable when applied to a realistic structure, i.e. that this model is indeed able to reproduce observations. As a first application, we consider the case of M33, a spiral galaxy at a distance of $0.7 \ {\rm Mpc}$ with an excellently measured rotation curve \citep{CP00} extending up to 13 times the disc half mass radius.

Before examining the fit to these data, we have to discuss preliminary how the galaxy is modelled. As usual, we split the total mass in two components, the stellar one and the dark matter halo. The surface brightness profile is well fitted by an exponential profile so that we assume an infinitesimally thin disc for the stellar component and compute the circular velocity as \citep{F70}\,:

\begin{equation}
v_{d}^2(R) = (2 G M_d/R_d) y^2 \left [ I_0(y) K_0(y) - I_1(y) K_1(y) \right ]
\label{eq: vcdiscnewt}
\end{equation}
where $y = R/2R_d$, $I_n(y)$ and $K_n(y)$ are the modified Bessel functions of order $n$ of the first and second kind, respectively, and the total disc mass is $M_d = \Upsilon_{\star} L_d$ with $L_d$ the total luminosity and $\Upsilon_{\star}$ the constant stellar mass\,-\,to\,-\,light ratio\footnote{We stress that  $\Upsilon_{\star}$ only refers to the stellar component and must not be confused with the total $M/L = M(R)/L(R)$ with $M(R) = M_{disk}(R) + M_{DM}(R)$ which is not constant.}. Photometric measurements allow to set $L_d = 4.2 \times 10^{9} \ {\rm L_{\odot}}$ and $R_d = 1.2 \ {\rm kpc}$ leaving $\Upsilon_{\star}$ as only parameter.
\begin{figure}
\centering
\includegraphics[width=7.5cm]{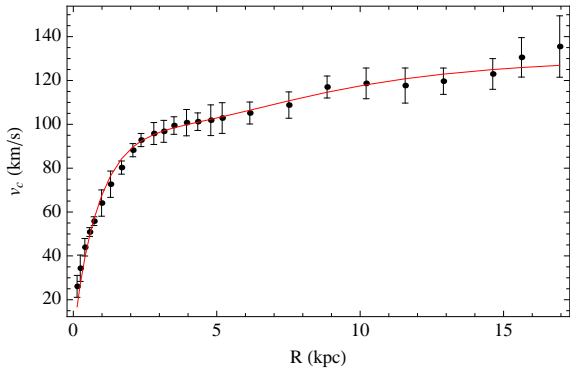}
\caption{Best fit rotation curve superimposed to M33 data.}
\label{fig: bfrc}
\end{figure}

Next, adopting the nonextensive context to model the dark halo density distribution the question arises whether this approach is self consistent. Indeed, the derivation was performed with respect to a system of collisionless self\,-\,gravitating particles, implicitly assuming a single component environment. Actually, we consider now a two component structure and, moreover, the stellar system is described by a model which is not the equilibrium configurations of a given statistics\footnote{We note that, actually, application of our formalism to the disc is not possible since baryons are subjected not only to gravitational interactions, but also to complicated phenomena describing the formation and evolution of stars. Taking care of all these ingredients in a statistical description is a daunting task that, as far as we know, has still to be addressed.}. However, we may assume that our approach applies to the proto\,-\,galactic systems providing the equilibrium configuration for the dark matter particles before the onset of baryonic collapse. Then it is still possible using the resulting model for the present day halo, provided that the disc onset has not significantly modified the starting configuration of DM particles. This should be a good approximation everywhere unless in the very inner regions where stars give the main contribution. However, the error induced on the total circular velocity should be negligible in case that these inner regions are stars dominated.

With this caveat in mind, we can fit the rotation curve of M33 in order to check the viability of the nonextensive halo model and constrain the stellar $M/L$ ratio and the halo parameters. To this end, we adopt a Bayesian approach and maximize the likelihood function ${\cal{L}}({\bf p}) \propto \exp{[- \chi^2({\bf p})/2 ]}$ with ${\bf p} = (\Upsilon_{\star}, \eta, {\cal{K}}, \alpha_0, r_0, \sigma_0)$ and

\begin{displaymath}
\chi^2({\bf p}) = \sum_{i = 1}^{{\cal{N}}_{obs}}{\left [ \frac{v_c^{obs}(R_i) - v_c^{th}(R_i, {\bf p})}{\varepsilon_i} \right ]^2} \ ,
\end{displaymath}
where $v_c^{obs}(R_i)$ and $v_c^{th}(R_i, {\bf p})$ are the observed (with error $\varepsilon_i$) and theoretically predicted circular velocity at $R_i$ and the sum is over the ${\cal{N}}_{obs}$ points. In order to efficiently sample the six dimensional parameter space, we use a Monte Carlo Markov chain algorithm running two chains with 100000 points each and checking convergence through the Gelmann\,-\,Rubin criterium \citep{GR92}. The marginalized constraints on the model parameters are then obtained by the histograms of the merged chains after cutting out the burn in phase and thinning to avoid spurious correlations.

\begin{table}
\begin{center}
\begin{tabular}{cccccc}
\hline
$x$ & $x_{BF}$ & $\langle x \rangle$ & $x_{med}$ & $68\%$ CL & $95\%$ CL \\
\hline \hline
~ & ~ & ~ & ~ & ~ & ~ \\
$\Upsilon_{\star}$ & 1.21 & 1.22 & 1.21 & (1.17, 1.26) & (1.12, 1.31) \\
~ & ~ & ~ & ~ & ~ & ~ \\
$\eta$ & 5.25 & 1.99 & 1.20 & (0.33, 2.86) & (0.13, 10.86) \\
~ & ~ & ~ & ~ & ~ & ~ \\
${\cal{K}}$ & 26.27 & 20.86 & 13.00 & (6.90, 39.93) & (3.25, 74.22) \\
~ & ~ & ~ & ~ & ~ & ~ \\
$r_0$ & 13.08 & 15.06 & 13.24 & (8.92, 21.92) & (6.40, 32.13) \\
~ & ~ & ~ & ~ & ~ & ~ \\
$\sigma_0$ & 76.3 & 101.5 & 96.1 & (79.9, 128.3) & (53.7, 146.6) \\
~ & ~ & ~ & ~ & ~ & ~ \\
\hline
\end{tabular}
\end{center}
\caption{Constraints on the model parameters from the fit to the M33 rotation curve. Columns are as follows\,: 1. parameter id; 2. best fit; 3., 4. mean and median from the marginalized likelihood; 5., 6. $68$ and $95\%$ confidence ranges. Note that we do not report constraints on $\alpha_0$ for the motivations discussed in text. Finally, $(\Upsilon_{\star}, r_0, \sigma_0)$ are in units of $(M_{\odot}/L_{\odot}, {\rm kpc}, {\rm km/s})$ respectively, while $(\eta, {\cal{K}})$ are dimensionless numbers.}
\label{tab: fitres}
\end{table}

The best fit parameters turns out to be\,:

\begin{displaymath}
\Upsilon_{\star} = 1.21 \ \ , \ \ \eta = 5.25 \ \ , \ \ {\cal{K}} = 26.27 \ \ ,
\end{displaymath}

\begin{displaymath}
\alpha_{0} = 0.0 \ \ , \ \ r_0 = 13.08 \ {\rm kpc} \ \ , \ \ \sigma_0 = 76.4 \ {\rm km/s} \ \ ,
\end{displaymath}
giving an excellent fit to the data as shown in Fig.\,\ref{fig: bfrc} and also confirmed by the value of the reduced $\chi^2$ ($\chi^2/d.o.f. = 0.59$). Although such a low $\chi^2$ could also signal that the errors are overestimated (as it is possible given how they have been estimated), the very good agreement between the theoretical and observed curve and the lack of any systematic trend of the residuals allow us confidently arguing that the model is indeed a good representation of the data.

Table \ref{tab: fitres} provides the constraints on the model parameters from the rotation curve fit. Apparently, the confidence ranges are quite large for all the cases except the stellar $M/L$ ratio. Indeed, this is a consequence of the severe degeneracy among the halo parameters which could be solved by adding some physically motivated priors. Table \ref{tab: fitres} does not report constraints on $\alpha_0$. Actually, our code gets stuck near $\alpha_0 = 0$ and starts wandering around this point because of a local minimum. We have tried to avoid this technical difficulty by running shorter chains for fixed values of $\alpha_0$ finding out that the constraints on the other parameters are nearly unaffected and therefore concluding that it is not possible to constrain this quantity with the data at hand. Actually, such a result could be anticipated noting that all the quantities we have discussed up to now are essentially independent on this parameter unless one is able to probe the very inner regions $(x << 0.01)$ which is not possible observationally. Since we are here interested in probing the viability of our model rather than constraining its parameters, we will not try to narrow down the above confidence ranges, but we plan to do this for a larger galaxy sample in a forthcoming work.

\section{Conclusions}

Modelling galaxies is one of the classical topics of galactic dynamics. While empirical, observationally motivated and N\,-\,body simulation inspired, models are widely used, they nevertheless lack a derivation from an underlying theoretical approach and do not allow to understand the effect of the different phenomena shaping the mass density profile. As a classical alternative, one can rely on the analogy between stars in a galaxy and molecules in a gas to extend the methods of statistical mechanics to the study of galactic systems. In this framework, we made a step forward investigating which density profile turns out as equilibrium configuration in the context of Tsallis nonextensive statistics.

Under the assumption of spherical symmetry we demonstrated that power\,-\,law DFs can solve the generalized Boltzmann equation and, when coupled to the Poisson equation, generate self-consistent models, which can be used to describe dark matter haloes. Subsequently, we investigated in detail the properties of the underlying mass model showing that its attractive features provide an interesting alternative to empirical density laws. As a test case, we have also fitted with remarkable success the M33 rotation curve, providing a significant observational support to the theoretical context.

A key ingredient within our derivation was the assumption of a parameterized expression for the velocity dispersion term entering the power\,-\,law DF. Although the resulting model is well behaved both theoretically and observationally, it is nevertheless worth noticing that using Eq.(\ref{eq: sigmamodel}) rather than another functional expression is actually only a matter of taste. Indeed, as far as we know, there are no theoretical motivations which can drive the choice of this function, wherefore it is only possible to introduce a heuristic approach and checking a posteriori the viability of a given $\sigma(r)$ profile. Alternatively, one could assume an expression for $q(r)$ and then solve Eqs.(\ref{eq: qsigmaeqend})\,-\,(\ref{eq: rhoeqend}) to get both $\rho(r)$ and $\sigma(r)$ but the choice of a functional form for $q(r)$ is difficult since we can only argue that $q(r)$ should approach the value of 1 asymptotically. The question of the central value of $q(r)$ and how the asymptotic regime is approached is far to be understood, implying that this path seems to be affected by a large degree of arbitrariness.

Regarding the uncertainty which approach to follow, the present proposal must be viewed as a first reasonable step requiring furhter investigations of the observational viability. Indeed, the successful fit of the M33 rotation curve appears as strong evidence in favour of the nonextensive approach, but is just a first step. In continuation it has to be checked whether this result is confirmed by fitting a larger sample of galaxies, spanning a large mass and luminosity range. On the other hand, rotation curves are hardly measured in elliptical galaxies implying that for these systems one has to fit the velocity dispersion profile. Using both stars and, e.g., planetary nebulae allows to trace the line of sight velocity dispersion up to large radii, thus providing the possibility of both, checking whether the nonextensive model applies also for theses systems and constraining the parameters involved. Finally, proceeding to higher redshift, the combined fit to projected mass and aperture velocity dispersion in lens systems (such as the SLACS sample) should test not only the model, but offer also hints on the evolution of the underlying parameters.

If all the above tests can be performed successfully, we would have not only a dark halo model inspired by fundamental principles, but also a new recipe to construct, in a self consistent way, dynamical models for galaxies in view of further steps towards an understanding of the nature of one of the main components in our cosmos.

\section*{Acknowledgments}

VFC warmly thanks Nicola R. Napolitano for drawing his attention to the empirical formula (\ref{eq: sigmamodel}) for $\sigma(r)$ and Paolo Salucci for pointing at the M33 data as a test case. The authors also acknowledge the referee for his/her constructive comments which have helped to improve the paper.

\end{document}